Modified Entropy Measure for Detection of Association Rules Under Simpson's Paradox Context.

Murphy Choy
Cally Claire Ong
Michelle Cheong

**Abstract**

The rapid explosion in retail data calls for more effective and efficient discovery of association rules to develop relevant business strategies and rules.Unlike online shopping sites, most brick and mortar retail shops are located in geographically and demographically diverse areas. This diversity presents a new challenge to the classical association rule model which assumes a homogenous group of customers behaving differently. The focus of this paper is centered on the discovery of association rules that were hidden as a result of a geographical and demographically diverse data. We will introduce a novel measure which incorporates the entropy measure with modified weighting for the detection of association rules not detected by the standard measures due to Simpson's paradox. The proposed measure is evaluated using a real-word case study involving a major retailer of fashion good in the context of traditional brick and mortar setting.

**Introduction**

Rapid advances in data management technologies has enabled large corporations to effectively store transactional data for retail stores at affordable rate. The sucess of such data collection exercises has spurned corporations that have access to huge data to attempt extracting important information from these sources to gain competitive advantage. The most common approach for analyzing huge transactional database is to apply market basket analysis, otherwise known as association rule mining. The technique allows for the discovery of customer purchasing behavior by examining the common occurrence of items in the transactional database. These discoveries can lead to interesting item bundling, service bundling as well as layout of retail stores.

The association model was first introduced by Agrawal et. Al (1993). Ever since then, there have been extensive studies of the model which resulted in a proliferation of the model. The extensions that have been researched in recent years, includes algorithm improvements (Brin et. Al., 1997; Han et. Al., 2000; Liu et. Al., 2002; Rastogi and Shim, 2002), fuzzy rules enhancements (Ishibuchi et. Al., 2001), multi-level and generalized rules (Clementini and Koperski, 2000), spatial rules, interesting rules and temporal association rules (Ale et. Al., 2000; Lee et. Al., 2001). The simplicity of association rules and ease of understanding, there have been many successful business applications in areas such as finance, telecommunication, retailing, and web design analysis. The most commonly cited examples are mainly found in retail industry.

Any major retailers is likely to have stores or branches in different geographical locations with a diverse demographic population. For such company with multiple stores, any purchasing behavior can be influenced by factors such as marketing, sales, service, and operation strategies at the company, local, and store levels. There are two challenges to the use of existing methods in any multi-store context. The first challenge comes from the dimension of time. As purchasing patterns can be seasonal even within a day, they can post difficulties in establishing the suitability and relevance of the rules without being able to determine the time period of relevance (Ale et. Al., 2000; Lee et. Al., 2001). This problem is addressed by temporal association models which are designed to find point-in-time behaviors or to search for time-invariant rules. For the models that are designed to detect point-in-time patterns, the period of relevance is incorporated into the computation of the support value. In this paper, we will not consider time as a factor due to the continuation nature of fashion goods where an updated design will replace an older design.

The second problem is found in problems involving the retailers having stores in diverse locations which differs both geographically as well as demographically (Chen et. Al., 2005). This new dimension can have a serious effects on the kind of product purchasing behavior. For example, a place with a well-to-do neighborhood is more likely to have purchases of more expensive categories of goods and brands compared to one found near rental housing area dominated by poorer individuals. At a location where most households have kids, the likelihood of toy and baby toiletries purchases are more likely than at a location more known as a retirement village. This makes it important to identify the rules which are location specific and can be influential in that area. At the same time, we should not penalize rules which are common in many areas. While this problem is considered to be related to spatial association rules, the focus is different as they focuses on the impact of geographical information rather than whether the rules are useful and whether they can be applied to multiple locations. The spread of the stores will be the focus of the paper where we will attempt identifying transactions which are important in different areas which are not identified by the traditional algorithm due to aggregation.

The paper is organized as follows. We will define the market basket problem in Section 2. In Section 3, we will propose a modified support measure and compare it against a panel of measures widely used in market basket. In Section 4, we compare the rules selected from the use of each and evaluate them in the context of a real retailer operating in a multi-store environment. The conclusion and future direction will be discussed in Section 5.

**Association Model**

Given two products, X and Y, an association rule takes the form of X => Y which implies a purchase pattern where a purchase of X is likely to lead to a purchase of Y. Support and confidence are two measures generally employed to select the most relevant association rules. Support measures the frequency of transactional records in the database having both X and Y, while confidence measures the accuracy of the rule which is defined as the percentage of transactions with both X and Y to the transactions with X. These measures are used in the A-priori algorithm (Aggarwal at. Al., 1993) which is the most commonly implemented algorithm for association rule mining. Let us consider the problem in a mathematical form below.

Let us assume a transactional database D that contains transactional records from multiple stores. We need to extract the association rules from the database. Let $I = (I_1, I_2, ..., I_r)$ be the set of product items that exists in D, where $I_k (1 \leq k \leq r)$ represent the $k^{th}$ item in D. The support for $X$ that is composed of the relation $I_i \rightarrow I_j$ is represented by $support(X, D)$, is the percentage of transactions containing X in database D,

$$support(X, D) = P(I_i, I_j)$$

For a specified support threshold $\rho_s$, X is a frequent itemset if

$$support(X, D) \geq \rho_s$$

The definitions of the support and the frequent itemset given above are used in the evaluation of traditional association rules. As seen in the formulation, the store information is not incorporated in the formula of the support of an item set. There have been extensive research in the measures used in market basket analysis (Tan et. Al., 2004). We will hereby list some of more common measures used in market basket analysis.

The Interest measure is commonly used in market basket analysis to evaluate the "interestingness" of a rule. This measure is used as a measure of deviation from statistical independence. However, it is sensitive to the support of the items $I_i$ or $I_j$. It is defined as,

$$Interest(X,D) = \frac{P(I_i, I_j)}{P(I_i)P(I_j)}$$

The Cosine Similarity Index is a simple index that is used to evaluate the association level between two items. It is closely related to the Interest Measure and is the geometric mean between interest measure and the support measure. The measure for pairs of items is also equivalent to the cosine measure commonly used as similarity measure for vector-space models. It is defined as,

$$Cosine(X,D) = \frac{P(I_i, I_j)}{\sqrt{P(I_i)P(I_j)}}$$

The Jaccard measure is used extensively in information retrieval to measure the similarity between documents. The Jaccard measures estimates the similarity between sample sets of the items. It is defined as,

$$Jaccard(X,D) = \frac{P(I_i, I_j)}{P(I_i) + P(I_j) - P(I_i, I_j)}$$

The Entropy measure is related to the variance of a probability distribution. The entropy of a flat distribution is large, whereas the entropy of a very skewed distribution is small. Given any two strongly associated items, the amount of reduction in entropy is high. The Entropy measure is defined as,

$$Entropy(X,D) = -P(I_i, I_j)\log(P(I_i, I_j)) - (1 - P(I_i, I_j))\log(1 - P(I_i, I_j))$$

In the next section, we will discuss about the Simpson's paradox in market basket analysis and how it affects the traditional market basket analysis. We will discuss about the use of the various measures and what are suitable cut off values for them.

**Simpson's Paradox and the Modified Entropy Measure**

In most data analysis, the association rule is usually deployed on the entire data set collected from different stores to extract key rules. However rules extracted this way may not be applicable in certain stores due to the geographical location or demographic factors. This nullification or even reversal of association is commonly known in Statistics as the Simpson's Paradox or Yules-Simpson's effect. One main reason why this effect is particularly obvious in market basket analysis is due mainly to the way the association rules are constructed. Most of the rules depends on confidence and support which are computed across stores. For example, good A is popular among the population but good B is usually bought with good A for a small group of people for a cluster of stores.

Let P(A) = 0.4 and P(B) = 0.01 and P(A and B) = 0.005

Support(A and B) = 0.005
Interest = 1.25

From the support and interest value, we can see that it is not a strong association nor an interesting one. However, when we move down to the store cluster, the effect is very different.

Let P(A|X) = 0.3 and P(B|X) = 0.1 and P(A and B|X) = 0.07

Support(A and B|X) = 0.07
Interest = 2.33

From the new cluster level support and interest value, we can see that association has improved in terms of support and is considered an interesting one. The common approach to the problem is to adopt a small support threshold which has a side effect of an explosion in the number of rules generated. Selection of proper confidence threshold is even more complicated and subtle. Given that the measures have few commonality among them, it is very difficult to compare the values. The simplest measure which can be compared with the classic support measure is the entropy measure. The entropy is a measure of the information value that is contained in the data. The entropy measure used in this case is the binary entropy measure. A comparison of the support value and entropy values are listed below.

| Support(%) | Entropy |
|---|---|
| 1.00% | 8.08% |
| 2.00% | 14.14% |
| 3.00% | 19.44% |
| 4.00% | 24.23% |
| 5.00% | 28.64% |
| 10.00% | 46.90% |
| 20.00% | 72.19% |
| 30.00% | 88.13% |
| 40.00% | 97.10% |
| 50.00% | 100.00% |

Table 1: Support and Entropy measure comparison

One of the key problems with the entropy measure is the strong bias for small entropy values. This presents a problem for the rules as small entropy values tend to favor rules which has very small support. As the minimum support is usually set as 1%, the equivalence in entropy value will be around 8%. However, too high a support level tend to lead to rules which are uninteresting. Given this, the rule filtering will be capped with a maximum level of 10% which has an entropy equivalence of 46.90%.

For Interest Measure, the common value used to determine whether rules are interesting is the cutoff of 3. Given the relation and similarities between Interest measure, Cosine measure and Jaccard measure, we will use a value of 5% for Cosine and 1% for Jaccard. However, for all the measures discussed so far, they do not incorporate the store level information. The author propose a modified entropy measure which combines a store measure with entropy function. The proposed function is as shown below,

$$SL\ Entropy(X,D,X,Y) = \log(\frac{Y}{X})(-P(I_i, I_j)\log(P(I_i, I_j)) - (1 - P(I_i, I_j))\log(1 - P(I_i, I_j)))$$

Where Y is the total number of stores and X being the number of stores where the rules is found.

The new measure incorporated the store level distribution information together with the entropy measure. The measure increases the entropy values by inflating it with a measure that identifies the number of stores which contributed to the rules. If the rules is found in small clusters, the entropy will be improved to indicate a good support for it in those clusters. The measure also reward single store rules which have very entropy values that deserves a second look. The SL Entropy measure is similar in construct to the TD*IDF measure widely used in text mining. Both measures combine a frequency measure with an inverse coverage measure to establish appropriate measurement value for useful and interesting information. In the next section, we will test the measure in a fashion good retail context with a myriad of stores.

**Real World Case Study: Fashion Good Retailer**

We evaluate our new association rule scoping measure using a multiple store transactional dataset provided by a fashion good retailer. The fashion good retailer is one of the leading fashion good provider in the country and has stores in various locations with different demographics. There are three different types of stores which are located at different malls and area. The first type of stores are located in super luxury malls patronized by the richer populace. The second type of stores are located in boutique shops dedicated to the brands. The final type of stores are located in departmental stores as display counters.

We will first analyze the rules generated by the classic association model for each of the store type and compare them to the overall rules generated. Below is the preliminary result.

| Transaction | T1 | T2 | T3 | Raw |
|---|---|---|---|---|
| A1=>A2 | 0 | 1 | 1 | 1 |
| A1=>A3 | 0 | 0 | 1 | 0 |
| A1=>A4 | 0 | 1 | 0 | 0 |
| A1=>A5 | 0 | 1 | 0 | 0 |
| A1=>A6 | 0 | 1 | 0 | 0 |
| A1=>A7 | 0 | 1 | 0 | 0 |
| A2=>A3 | 0 | 1 | 0 | 0 |
| A2=>A4 | 0 | 1 | 0 | 0 |
| A2=>A5 | 0 | 1 | 1 | 1 |
| A2=>A6 | 0 | 1 | 0 | 0 |
| A2=>A7 | 0 | 1 | 0 | 0 |
| A2=>A8 | 0 | 1 | 0 | 0 |
| A3=>A6 | 0 | 1 | 0 | 0 |
| A5=>A7 | 0 | 1 | 0 | 0 |
| A6=>A5 | 0 | 1 | 0 | 0 |
| A6=>A7 | 0 | 1 | 0 | 0 |
| A4=>A5 | 0 | 1 | 0 | 0 |
| A4=>A6 | 0 | 1 | 0 | 0 |
| A8=>A9 | 0 | 1 | 0 | 0 |
| Sub Total | 0 | 18 | 3 | 2 |
| A10=>A11 | 1 | 0 | 0 | 0 |
| A11=>A12 | 1 | 0 | 0 | 0 |
| A11=>A13 | 1 | 0 | 0 | 1 |
| A11=>A14 | 1 | 0 | 0 | 1 |
| A11=>A15 | 1 | 0 | 0 | 0 |
| A11=>A16 | 1 | 0 | 0 | 0 |
| A13=>A14 | 1 | 0 | 0 | 1 |
| A13=>A15 | 1 | 0 | 0 | 0 |
| A13=>A16 | 1 | 0 | 0 | 0 |
| A14=>A15 | 1 | 0 | 0 | 0 |
| A14=>A16 | 1 | 0 | 0 | 0 |
| A15=>A16 | 1 | 0 | 0 | 0 |
| A16=>A17 | 1 | 0 | 0 | 0 |
| Sub Total | 13 | 0 | 0 | 3 |
| Total | 13 | 18 | 3 | 5 |

**Table 2: Store Level comparison**

From the results in table 2, we can observe that the type 1 and 2 stores are radically different in terms of the relations and that the classic model on the entire database indicated by raw has only detected 5 rules using the classic measures. This is poor as the bulk of the rules found in the three types of stores make up a total of 32 rules of which only 5 are detected at overall level. That is less than 50% of the rules discovered. Let us compare the performances of the other measures as well as the new SL Entropy measure.

| Transaction | T1 | T2 | T3 | Raw | IS | Jaccard | Entropy | SL Entropy |
|---|---|---|---|---|---|---|---|---|
| A1=>A2 | 0 | 1 | 1 | 1 | 1 | 1 | 1 | 1 |
| A1=>A3 | 0 | 0 | 1 | 0 | 1 | 1 | 1 | 1 |
| A1=>A4 | 0 | 1 | 0 | 0 | 0 | 0 | 0 | 1 |
| A1=>A5 | 0 | 1 | 0 | 0 | 1 | 1 | 1 | 1 |
| A1=>A6 | 0 | 1 | 0 | 0 | 0 | 1 | 1 | 1 |
| A1=>A7 | 0 | 1 | 0 | 0 | 0 | 1 | 1 | 1 |
| A2=>A3 | 0 | 1 | 0 | 0 | 1 | 1 | 1 | 1 |
| A2=>A4 | 0 | 1 | 0 | 0 | 0 | 1 | 0 | 1 |
| A2=>A5 | 0 | 1 | 1 | 1 | 1 | 1 | 1 | 1 |
| A2=>A6 | 0 | 1 | 0 | 0 | 1 | 1 | 1 | 1 |
| A2=>A7 | 0 | 1 | 0 | 0 | 1 | 1 | 1 | 1 |
| A2=>A8 | 0 | 1 | 0 | 0 | 0 | 0 | 0 | 0 |
| A3=>A6 | 0 | 1 | 0 | 0 | 0 | 0 | 0 | 0 |
| A5=>A7 | 0 | 1 | 0 | 0 | 0 | 1 | 1 | 1 |
| A6=>A5 | 0 | 1 | 0 | 0 | 1 | 1 | 1 | 1 |
| A6=>A7 | 0 | 1 | 0 | 0 | 0 | 1 | 1 | 1 |
| A4=>A5 | 0 | 1 | 0 | 0 | 0 | 0 | 0 | 0 |
| A4=>A6 | 0 | 1 | 0 | 0 | 1 | 1 | 0 | 1 |
| A8=>A9 | 0 | 1 | 0 | 0 | 0 | 0 | 0 | 0 |
| Sub Total | 0 | 18 | 3 | 2 | 9 | 14 | 12 | 15 |
| A10=>A11 | 1 | 0 | 0 | 0 | 1 | 1 | 0 | 1 |
| A11=>A12 | 1 | 0 | 0 | 0 | 1 | 1 | 0 | 1 |
| A11=>A13 | 1 | 0 | 0 | 1 | 1 | 1 | 1 | 1 |
| A11=>A14 | 1 | 0 | 0 | 1 | 1 | 1 | 1 | 1 |
| A11=>A15 | 1 | 0 | 0 | 0 | 1 | 1 | 1 | 1 |
| A11=>A16 | 1 | 0 | 0 | 0 | 0 | 0 | 1 | 1 |
| A13=>A14 | 1 | 0 | 0 | 1 | 1 | 1 | 1 | 1 |
| A13=>A15 | 1 | 0 | 0 | 0 | 0 | 1 | 0 | 1 |
| A13=>A16 | 1 | 0 | 0 | 0 | 0 | 0 | 1 | 1 |
| A14=>A15 | 1 | 0 | 0 | 0 | 1 | 1 | 1 | 1 |
| A14=>A16 | 1 | 0 | 0 | 0 | 1 | 1 | 1 | 1 |
| A15=>A16 | 1 | 0 | 0 | 0 | 0 | 0 | 1 | 1 |
| A16=>A17 | 1 | 0 | 0 | 0 | 0 | 0 | 1 | 1 |
| Sub Total | 13 | 0 | 0 | 3 | 8 | 9 | 10 | 13 |
| Total | 13 | 18 | 3 | 5 | 17 | 23 | 22 | 28 |

**Table 2: New measure and alternate measure comparison**

From the new results in table 2, we can observe that the Entropy, Cosine and Jaccard measures have performed much better than the traditional support measure in detecting rules which are found at store level. Entropy achieved a detection rate of 68.75%, Jaccard achieved 71.88% while cosine achieved 53.12%. The new SL Entropy achieved 87.5% detection rate which is far above the others. The new measure is much more effective than the other measures as well as the traditional measure.

## Conclusion

One critical challenge for store level association rule generation is the need to analyze data at overall level while compensating for the loss of granularity. We have proposed a novel measure of association rule for detection of rules which are at store level or dispersed across location. The new measure provides enhanced capabilities to retail companies to explore association rules which are based on clusters' behavior. We have evaluated the proposed measure in a real-world case study that focus on fashion good retail with brick and mortar stores. The measure was able to identify the rules identified at the store type level while analyzing the data at an overall level.

For future work, we plan to extend the measure beyond a single factor. Only a handful of the measures discussed can be generalized to multi-level interactions. Analyzing such multi-layered associations is much more complicated due to the exponential increase in the number of factors to be considered. The increased dimensions will also cause the number of relations that is needed to be increased to a level that allows for proper analysis. A good measure must be able to identify the associations among the factors from the overall data. Careful study is needed to establish the most appropriate measure for this purpose.